\documentclass[preprint,3p,12pt]{elsarticle}

\input{Qcircuit}
\usepackage{appendix,amsmath,amssymb,bbm,graphicx,nccmath,subfigure,stmaryrd}
\biboptions{sort,compress}

\newcommand{\mat}[2]{\ensuremath{\left( \begin{array}{#1} #2 \end{array} \right)}}

\newcounter{bla}

\begin{document}

\begin{frontmatter}

\title{\emph{OptQC}: An optimised parallel quantum compiler}
\author{T. Loke, J. B. Wang \footnote{corresponding author: jingbo.wang@uwa.edu.au} and Y.H. Chen}
\address{School of Physics, The University of Western Australia, 6009 Perth, Australia}

\begin{abstract}

The software package \emph{Qcompiler} \cite{chen_qcompiler:_2013} provides a general quantum compilation framework, which maps any given unitary operation into a quantum circuit consisting of a sequential set of elementary quantum gates.  In this paper, we present an extended software \emph{OptQC}, which finds permutation matrices $P$ and $Q$ for a given unitary matrix $U$ such that the number of gates in the quantum circuit of $U = Q^TP^TU'PQ$ is significantly reduced, where $U'$ is equivalent to $U$ up to a permutation and the quantum circuit implementation of each matrix component is considered separately.  We extend further this software package to make use of high-performance computers with a multiprocessor architecture using MPI.  We demonstrate its effectiveness in reducing the total number of quantum gates required for various unitary operators.

\vspace{2cm}

\noindent\textbf{Program summary}
\vspace{6pt}
\noindent \emph{Program title}: \emph{OptQC} \\
\vspace{6pt}
\noindent \emph{Catalogue identifier}: \\
\vspace{6pt}
\noindent \emph{Program summary}: URL: http://cpc.cs.qub.ac.uk/summaries/  \\
\vspace{6pt}
\noindent \emph{Program obtainable from}:  CPC Program Library, Queens University, Belfast, N. Ireland \\
\vspace{6pt}
\noindent \emph{Licensing provisions}: Standard CPC licence, http://cpc.cs.qub.ac.uk/licence/licence.html \\
\vspace{6pt}
\noindent \emph{Distribution format}: tar.gz \\
\vspace{6pt}
\noindent \emph{Programming language}: Fortran, MPI \\
\vspace{6pt}
\noindent \emph{Computer}: Any computer with Fortran compiler and MPI library. \\
\vspace{6pt}
\noindent \emph{Operating system}:  Linux \\
\vspace{6pt}
\noindent \emph{Classification}: \\
\vspace{-2pt}
\noindent \emph{Nature of problem}:  It aims to minimise the number of quantum gates required to implement a given unitary operation. \\

\vspace{-6pt}
\noindent \emph{~Solution method}:  It utilises a threshold-based acceptance strategy for simulated annealing to select permutation matrices $P$ and $Q$ for a given unitary matrix $U$ such that the number of gates in the quantum circuit of $U = Q^TP^TU'PQ$ is minimised, where $U'$ is equivalent to $U$ up to a permutation.  The decomposition of a unitary operator is performed by recursively applying the cosine-sine decomposition. \\

\vspace{-6pt}
\noindent \emph{~Running time}: Running time increases with the size of the unitary matrix, as well as the prescribed maximum number of iterations for qubit permutation selection and the subsequent simulated annealing algorithm. Running time estimates are provided for each example in Section \ref{sec:res}.  All simulation results presented in this paper are obtained from running the program on the Fornax supercomputer managed by iVEC@UWA with Intel Xeon X5650 CPUs.\\

\end{abstract}

\end{frontmatter}

\section{Introduction}

Quantum computation aims to solve problems that are classically intractable by harnessing intricate quantum correlations between densely encoded states in quantum systems \cite{nielsen_quantum_2011}. A well-known example is Shor's algorithm for the factorisation of numbers \cite{shor_polynomial-time_1997, vandersypen_experimental_2001}. Quantum algorithms are designed to be implemented on quantum computers by means of a quantum circuit, which consists of qubits and quantum gates. It is therefore of vital importance to be able to obtain a quantum circuit representation for any given quantum algorithm (which is always described by a unitary matrix) in terms of an elementary set of quantum gates - this role is that of a quantum compiler.

Barenco \emph{et al} and Deutsch \emph{et al} \cite{barenco_elementary_1995, deutsch_universality_1995} proved that any arbitrarily complex unitary operation can be implemented by a quantum circuit involving only one- or two-qubit elementary quantum logic gates.  Earlier studies applied the standard triangularisation or QR-factorisation scheme with Givens rotations and Gray codes to map a quantum algorithm to a series of elementary gate operations \cite{barenco_elementary_1995, deutsch_universality_1995, cybenko_reducing_2001, nielsen_quantum_2011}.  Several research groups examined a more efficient and versatile scheme based on the cosine-sine decomposition was proposed and utilized \cite{tucci_rudimentary_1999, mottonen_quantum_2004, bergholm_quantum_2005, khan_synthesis_2006, manouchehri_quantum_2009}.  De Vos  \emph{et al}  \cite{de_vos_multiple-valued_2009, alexis_de_vos_reversible_2012} looked into another decomposition scheme, namely the Birkhoff decomposition, which was found to provide simpler quantum circuits for certain types of unitary matrices than the cosine-sine decomposition.  However, the Birkhoff decomposition does not work for general unitary matrices.  

More recently, Chen and Wang \cite{chen_qcompiler:_2013} developed a general quantum compiler package written in Fortran, entitled the \emph{Qcompiler}, which is based on the cosine-sine decomposition scheme and works for arbitrary unitary matrices.  The number of gates required to implement a general $2^n$-by-$2^n$ unitary matrix using the CSD method scales as O($4^n$) \cite{mottonen_quantum_2004, manouchehri_quantum_2009}. In other words, the number of gates scales exponentially with the number of qubits. Thus, in any practical application of the CSD method to decomposing matrices, it is of considerable interest to reduce the number of gates required as much as possible. 

In this work, we adopt the CSD method due to the reasons outlined above, and we split the unitary matrix $U$ into an equivalent sequence of unitaries with the aim of reducing the number of gates required to implement the entire sequence of unitaries. In general, this means writing $U$ as a sequence of $s$ unitaries, i.e. $U = U_s U_{s-1} \ldots U_1$. At first glance, this seems counterintuitive, since if we were to apply the CSD to each unitary, this would increase the scaling of the number of gates required to O($4^ns$), which is undesirable. However, we note that (1) certain $U_i$ can be decomposed more efficiently than CSD such as qubit permutation matrices; and (2) some matrices requires only a few gates when separately decomposed using the CSD method.

This paper is organised as follows. Section \ref{sec:oa} describes in detail our approach for reducing the number of gates required to implement any given unitary matrix $U$. Section \ref{sec:po} details our developed program, called \emph{OptQC}, that uses the methods described in section \ref{sec:oa} to reduce the number of gates required to implement any given unitary matrix. Some sample results using the program are given in section \ref{sec:res}, and then we discuss our conclusions and possible future work in section \ref{sec:cafw}.

\section{Our approach}
\label{sec:oa}

Suppose we are given an $m$-by-$m$ (where $m=2^n$) unitary matrix $U$. As mentioned above, we are interested in splitting $U$ into a sequence of unitaries with the aim of reducing the total number of gates required to implement the entire sequence. One means of splitting $U$ into an equivalent sequence of unitaries is by using permutation matrices. A permutation matrix is a square binary matrix that contains, in each row and column, precisely a single 1 with 0s everywhere else. For any permutation matrix $P$, its corresponding inverse is $P^{-1} = P^T$. For convenience, we also define an equivalent representation of permutations using lists - a permutation list $p$ (lowercase) is equivalent to the permutation matrix $P$ (uppercase) by the relation: 
\begin{equation}
\left( P \right)_{i,j} = \delta_{p\llbracket i \rrbracket, j} ,
\label{eqn:pltom}
\end{equation}
where $\delta$ is the Kronecker delta function, and $p\llbracket i \rrbracket$ denotes the $i$th list element of $p$. For example, the permutation list $p = \{2,1,4,3\}$ corresponds to the 4-by-4 permutation matrix:
\begin{equation*}
P = 
\mat{cccc}
{
0 & 1 & 0 & 0 \\
1 & 0 & 0 & 0 \\
0 & 0 & 0 & 1 \\
0 & 0 & 1 & 0
} .
\end{equation*}

Now define $ \mbox{CSD}(M) $ to be the number of gates required to implement the unitary matrix $M$ according to the CSD method. If we were to write $U$ as $U = P^T U' P$ (where $U'$ is equivalent to $U$ up to a permutation), then we find that $ \mbox{CSD}(U) \neq \mbox{CSD}(U') + \mbox{CSD}(P) + \mbox{CSD}(P^T) $ in general (note also that $\mbox{CSD}(P) \neq \mbox{CSD}(P^T)$). The general aim is thus to find a $P$ that minimises the total cost function $ \mbox{CSD}(U') + \mbox{CSD}(P) + \mbox{CSD}(P^T) $, with the obvious restriction that it has to be less than $ \mbox{CSD}(U)$. 

In our approach, we write $U$ as $U = Q^T P^T U' P Q$, where $P$ and $Q$ are both permutation matrices, and $U'$ is equivalent to $U$ up to a permutation. In general, $P$ is allowed to be any permutation matrix ($m!=(2^n)!$ permutations possible), but $Q$ is restricted to a class of permutation matrices that correspond to qubit permutations (only $n!$ permutations possible). The advantage of this approach is that qubit permutations can be easily implemented using a sequence of swap gates - so for a system with $n$ qubits, it requires at most $n-1$ swap gates to implement any qubit permutation. This also enables the program (in the parallel version) to start different threads at different points in the search space of $m!$ permutations by using different qubit permutations.

Let $ s_{num}(Q) $ be the number of swap gates required to implement a qubit permutation matrix $Q$. Note that $s_{num}(Q) = s_{num}(Q^T)$, since the reverse qubit permutation would just be the same swap gates applied in reverse order. The total cost function $c_{num}$ of implementing a given unitary $U = Q^T P^T U' P Q$ is then:
\begin{equation}
c_{num}(U) = \mbox{CSD}(U') + \mbox{CSD}(P) + \mbox{CSD}(P^T) + 2 s_{num}(Q) .
\end{equation}
To make the dependencies in this function clear, we write this as:
\begin{equation}
c_{num}(U,P,Q) = \mbox{CSD}(P Q U Q^T P^T) + \mbox{CSD}(P) + \mbox{CSD}(P^T) + 2 s_{num}(Q) ,
\label{eqn:costf}
\end{equation}
which is the function that we aim to minimise with respect to $P$ and $Q$.


\section{Program outline}
\label{sec:po}

We have developed a Fortran program, called \emph{OptQC}, which reads in a unitary matrix $U$, minimises the total cost function $ c_{num}(U,P,Q) $, and outputs a quantum circuit that implements $U$. A significant portion of this program is based on the CSD code provided by the LAPACK library \cite{anderson_lapack_1999}) and the recursive procedure implemented in \emph{Qcompiler}, developed by Chen and Wang \cite{chen_qcompiler:_2013}. As with \emph{Qcompiler}, the new \emph{OptQC} program has two different branches, one treating strictly real unitary (i.e. orthogonal) matrices, and another treating arbitrary complex unitary matrices, with the former generally providing a circuit that is half in size of the latter \cite{chen_qcompiler:_2013}. 



Note that the CSD procedure requires the round up of the matrix dimension to the closest power of two, i.e. the dimension used is
\begin{equation}
m' = 2^{\lceil {log}_2 m \rceil}.
\label{eqn:nd}
\end{equation}

\noindent The expanded unitary operator $\bar{U} $ is an $m'$-by-$m'$ matrix, where

\begin{equation}
\left( \bar{U} \right)_{i,j} = \left\{
\begin{array}{ll}
\left( U \right)_{i,j} & : i \leq m, j \leq m \\
\delta_{i,j} & : \mbox{otherwise}
\end{array}
\right.
\label{eqn:Uexpand}
\end{equation}

\noindent which we will subsequently treat as the unitary $ U $ to be optimised via permutations.

\begin{figure}[htp]
	\centering
	\includegraphics[scale=0.30]{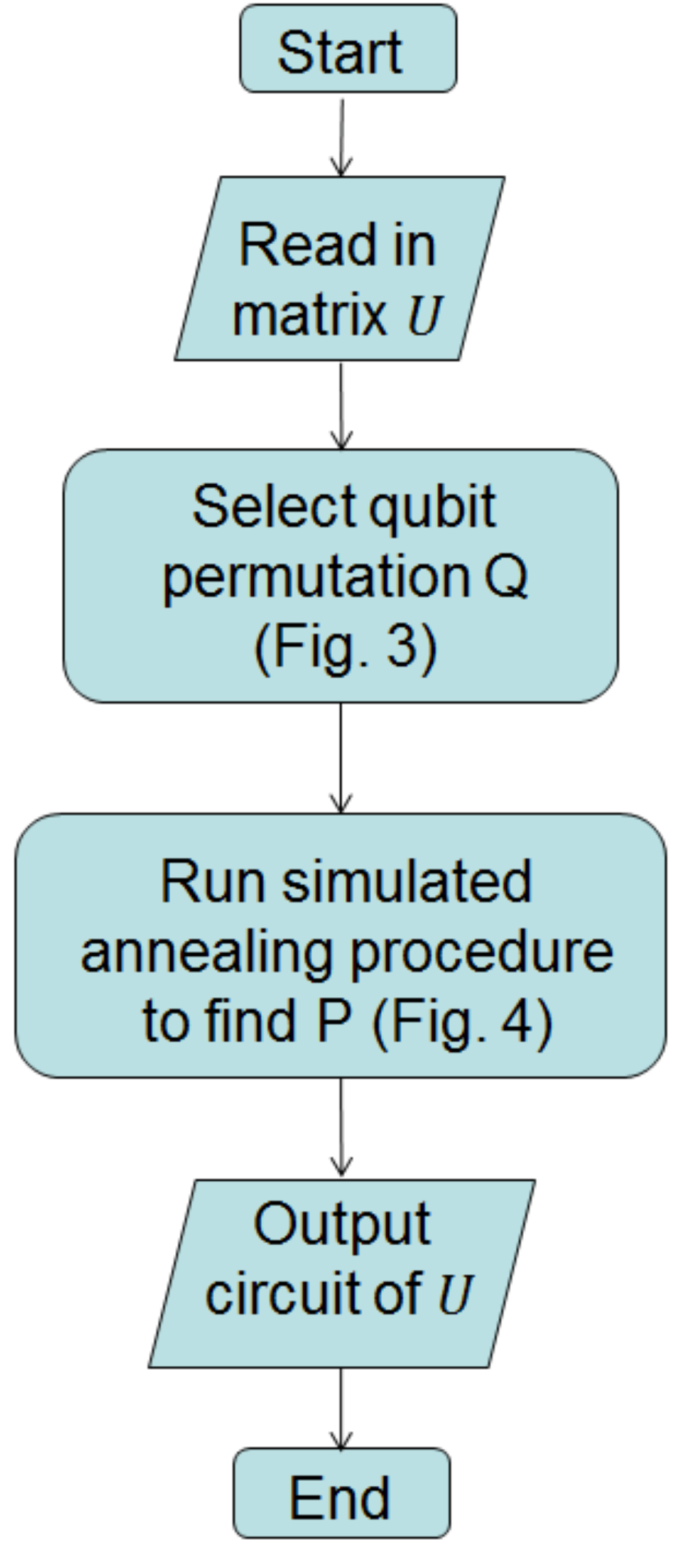}
	\caption{Flowchart overview of the serial version of \emph{OptQC}.}
	\label{fig:fchart}
\end{figure}

In the following subsections we describe the key procedures in \emph{OptQC}, depicted in Figure \ref{fig:fchart}, which serve to progressively reduce the total cost function $ c_{num}(U,P,Q) $. We first detail the serial version of the program, followed by an extension to a parallel architecture using MPI.

\subsection{Selection of qubit permutation}
\label{sec:initialq}

Qubit permutations are a class of permutations that are expressible in terms of a reordering of qubits, which can be efficiently implemented using swap gates that serve to interchange qubits. Recalling that $U$ is of dimensions $m$-by-$m$ (where $m=2^n$), this implies that there are only $n!$ qubit permutations possible for a given $U$.  A qubit permutation can be expressed as a list $q$ (lowercase) of length $n$, or as a permutation matrix $Q$ (uppercase) of dimensions $m$-by-$m$. A qubit permutation of length $n$ requires at most $n-1$ swap gates. 

\begin{figure}[htp]
\centering
\includegraphics[scale=0.20]{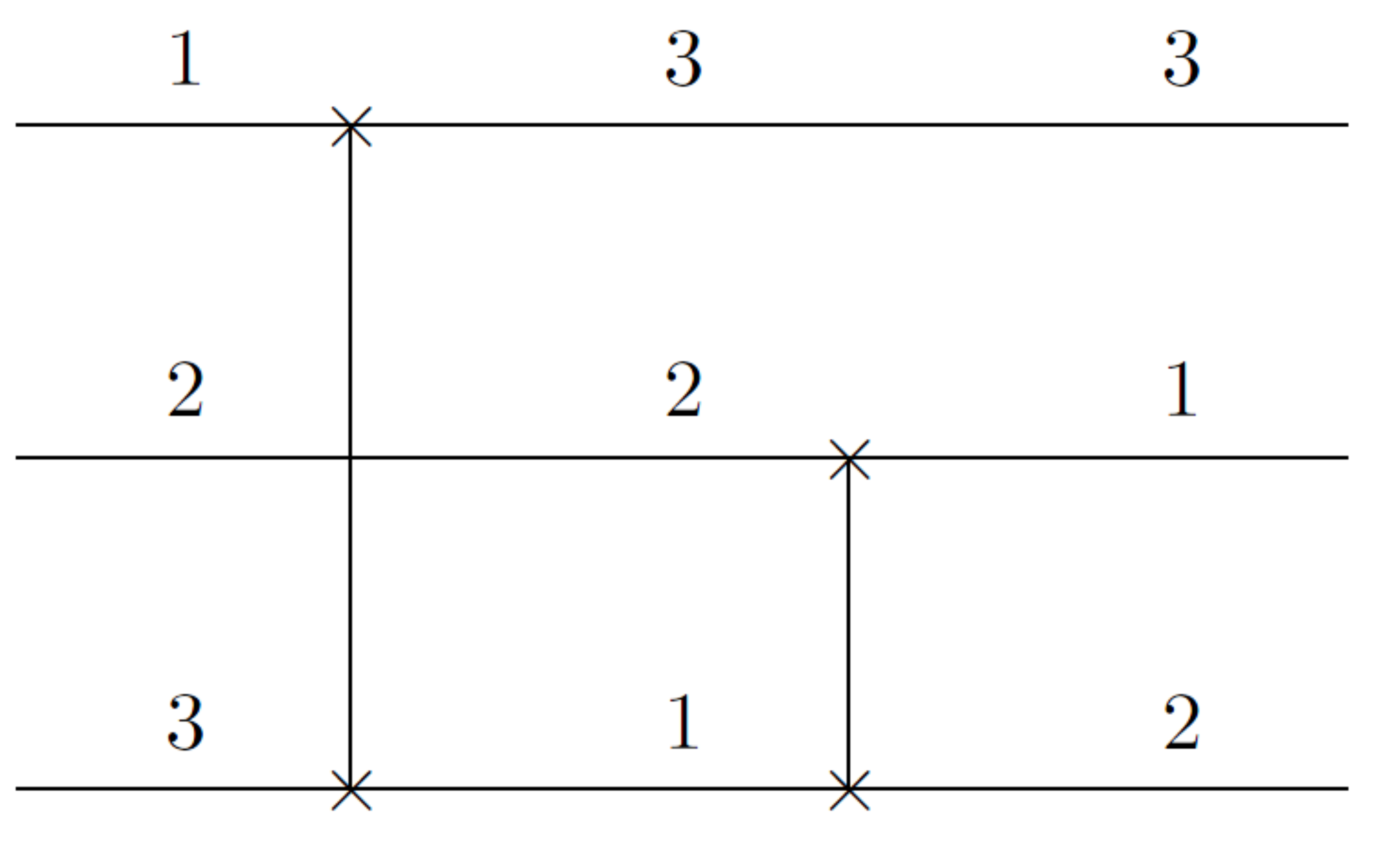}
\caption{Example implementation of qubit permutation $q = \{ 3, 1, 2 \}$ using swap gates.}
\label{fig:swapeg}
\end{figure}

The selection of the qubit permutation matrix $Q$ is done by varying $Q$ and computing the corresponding change in the cost function $c_{num}$, while holding $P$ constant as the identity matrix $I$. An example implementation of the $n=3$ qubit permutation $q = \{ 3, 1, 2 \}$ is shown in Figure \ref{fig:swapeg}. By considering how the basis states are mapped to each other by $q$, a regular permutation list $\bar{q}$ of length $m$ can be readily constructed from $q$, and then we use the relation between permutation lists and permutation matrices (see Eq.~(\ref{eqn:pltom})) to obtain $Q$ from $\bar{q}$.

\begin{figure}[htp]
\centering
\includegraphics[scale=0.30]{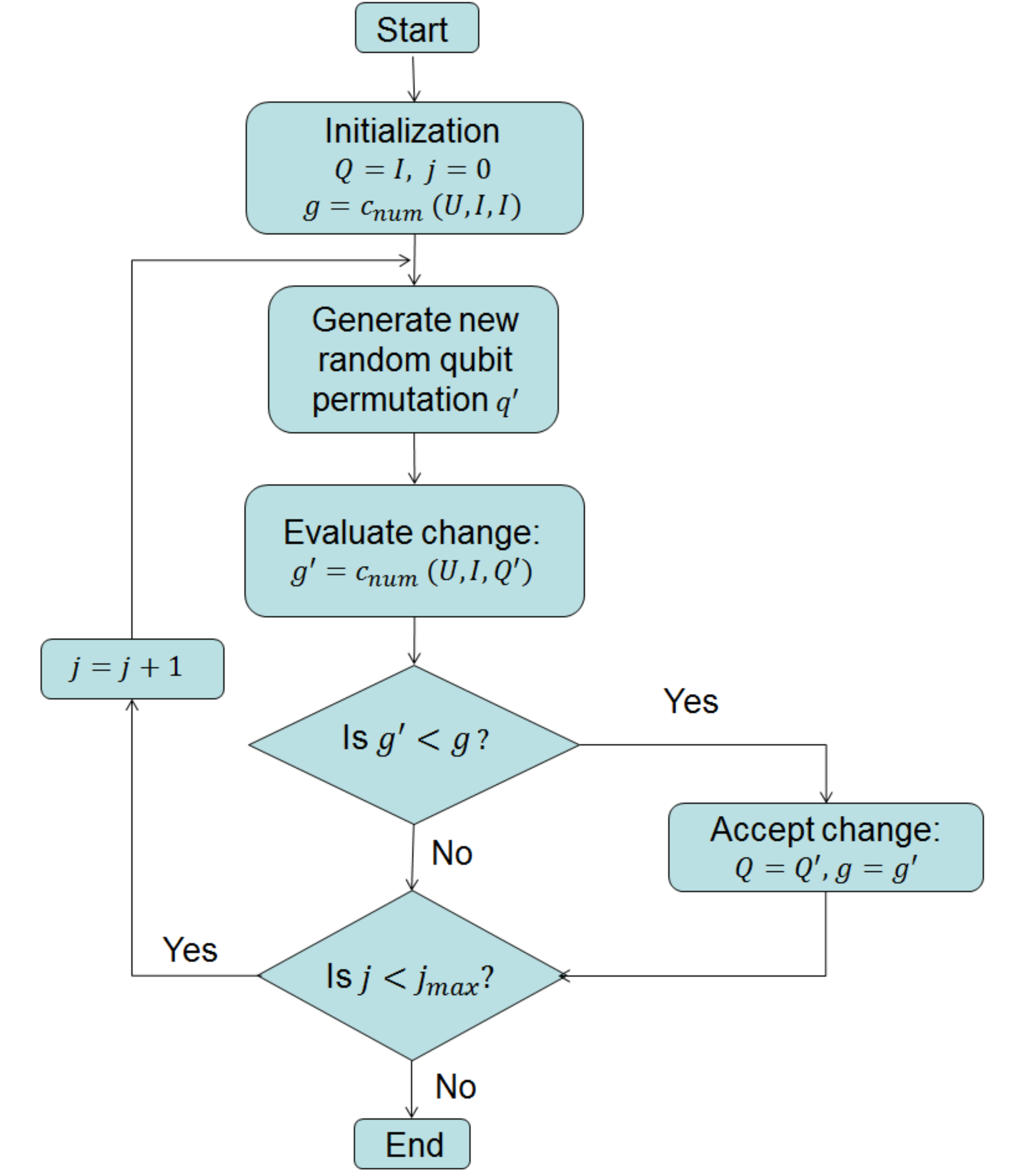}
\caption{Flowchart overview of the qubit selection procedure.}
\label{fig:qsfchart}
\end{figure}

We start the program with an identity qubit permutation $ q $, i.e. $ q[[i]] = i $ (corresponding to $ Q = I $), and compute the corresponding cost of implementation $c_{num}(U,I,Q)$.  Then, for some prescribed number of iterations $j_{max}$, we generate a random qubit permutation $ q' $ each time and compute the new cost as $c_{num}(U,I,Q')$. If the new cost is lower than the initial cost (recorded by $c_{num}(U,I,Q)$), the current qubit permutation $q$ is replaced by $q'$.  Figure \ref{fig:qsfchart} shows a flowchart overview of the qubit selection procedure. After this procedure, we have an optimised qubit permutation matrix $Q$, which will remain unchanged while we find the unrestricted permutation matrix $P$ in the next section through a simulated annealing process.   

\subsection{Simulated annealing}
\label{sec:simann}

Here, we aim to find an optimal permutation $p'$ such that $ c_{num}(U,P',Q) < c_{num}(U,P,Q) $ in the discrete search space of all $m!$ permutations. Given the massive size of the search space, use of a heuristic optimisation method is practically necessary. Simulated annealing is one such method for finding a minimum in a discrete search space.  In the OptQC program, we adopt a threshold acceptance based simulated annealing algorithm. There are three key components to the algorithm:

\begin{enumerate}
\item Cost function: the function to be minimised, i.e. $ c_{num}(U,P,Q) $.\\

\item Neighbourhood operator: the procedure that alters the current solution slightly by altering the current permutation $p$ to a slightly different permutation $p'$. Our neighbourhood operator acts to interchange any two \emph{random} positions in $p$ to form $p'$. \\

\item Threshold value: any 'bad' trades (increase in cost function) that are below some threshold value $\beta$ are accepted, otherwise they are rejected. We define the threshold value as $\beta(P,Q) = \mbox{min}\left(\lceil \alpha c_{num}(U,P,Q) \rceil,\lceil \alpha c_{num}(U,I,Q) \rceil\right)$, where $ 0 \leq \alpha < 1 $. As such, the threshold value is taken to be the proportion $\alpha$ of the current number of gates (with a fixed maximum value of the proportion $\alpha$ of the initial number of gates to ensure that $\beta(P,Q)$ cannot grow arbitrarily large).
\end{enumerate}

\begin{figure}[htp]
\centering
\includegraphics[scale=0.30]{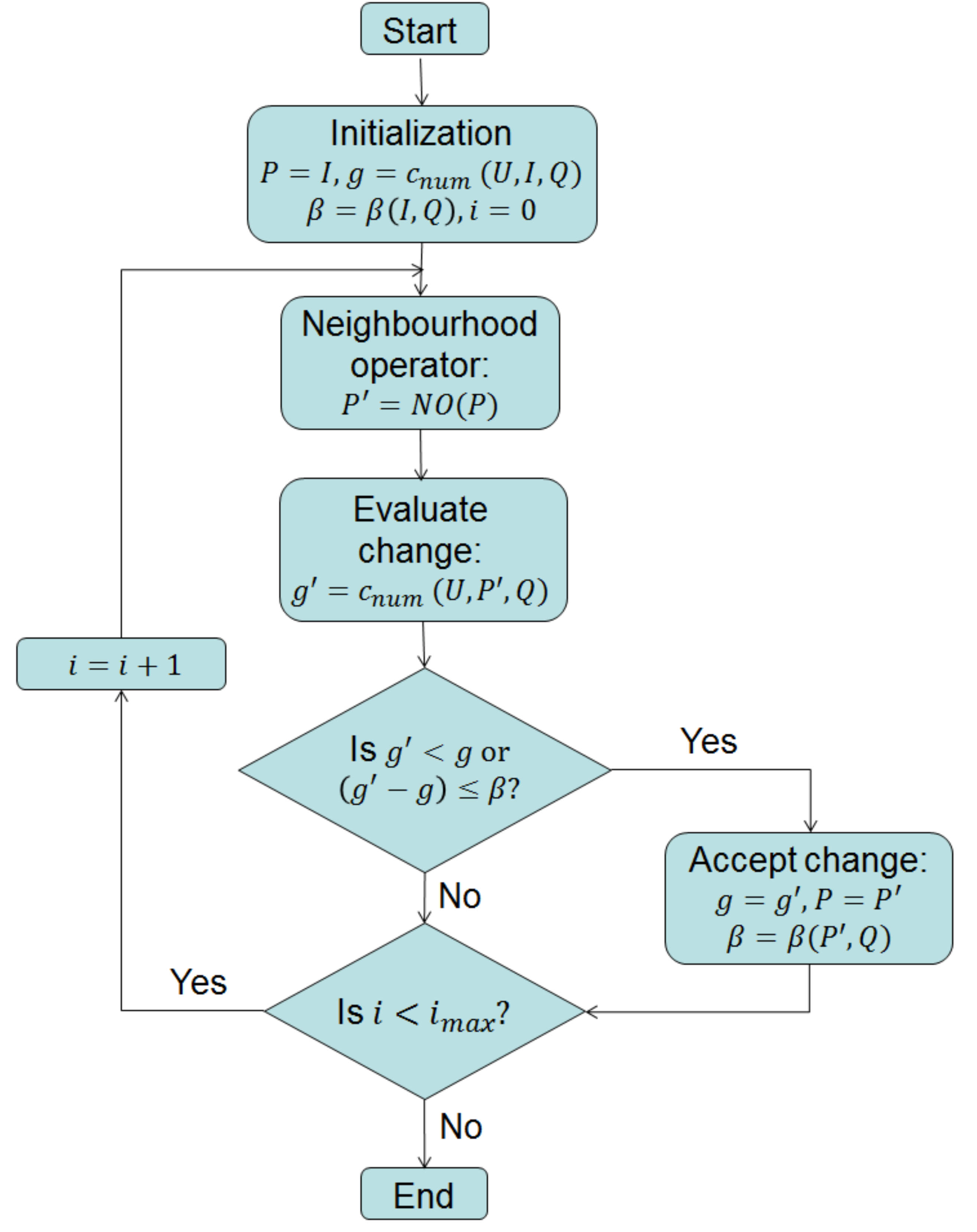}
\caption{Flowchart overview of the simulated annealing procedure.}
\label{fig:safchart}
\end{figure}

We start with $p$ as the identity permutation. By iterating the neighbourhood operator and evaluating the subsequent change in the number of gates, we accept the change in the permutation if it reduces the number of gates, or if the increase in the number of gates is below the threshold $\beta$. After some prescribed number of iterations $i_{max}$, we terminate the simulated annealing procedure, returning the permutation $p_{min}$ that provides the minimum number of gates. Figure \ref{fig:safchart} shows a flowchart overview of the simulated annealing procedure. Note that $p_{min}$ is not necessarily the permutation $p$ at the end of $i_{max}$ iterations - rather, we keep track of $p_{min}$ separately during the procedure.

\subsection{Gate reduction procedure}

\begin{figure}[htp]
\centering
\subfigure[Circuit before reduction procedure]{\label{fig:bred}\includegraphics[scale=0.30]{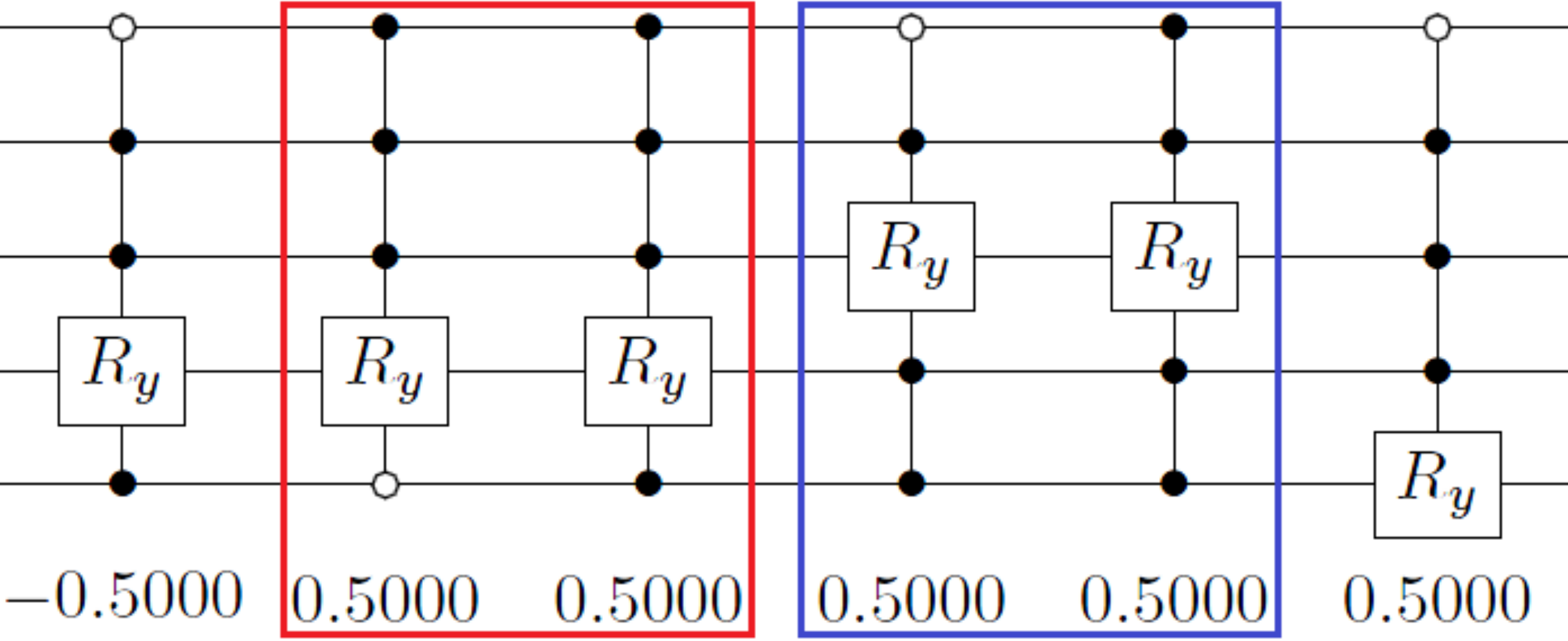}\qquad}
\subfigure[Circuit after reduction procedure]{\label{fig:rred}\includegraphics[scale=0.30]{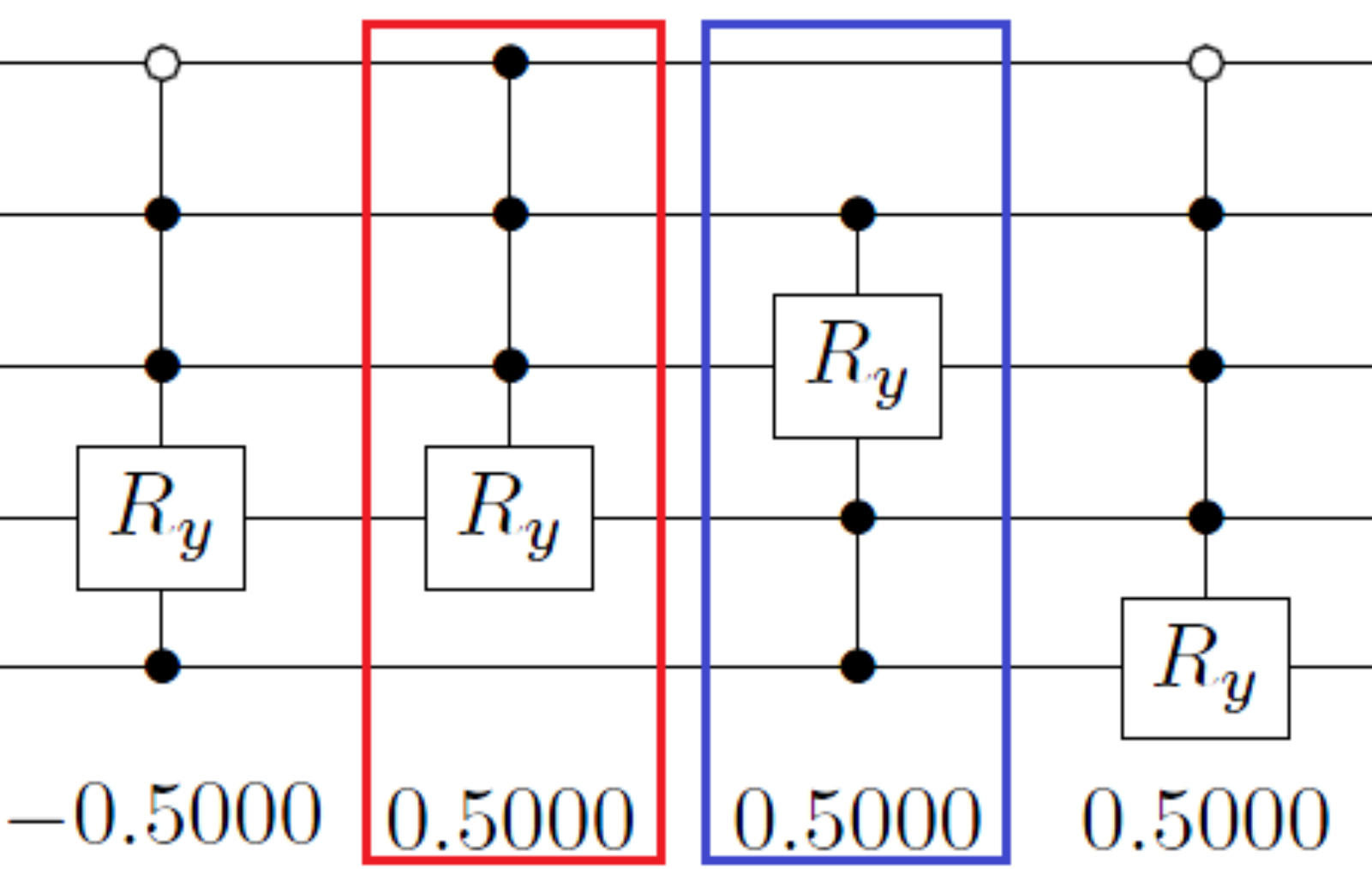}}
\caption{Example of applying the reduction procedure to a quantum circuit.}
\label{fig:red}
\end{figure}

Here, we focus on reducing the number of gates in some prescribed quantum circuit by combining 'similar' gates. In a quantum circuit, we can combine CUGs (controlled unitary gates) that apply the same unitary operation $U_{op}$ to the same qubit, with all but one of the conditionals of the CUGs being the same. This reduction process is carried out after every application of the CSD method to a matrix - in particular, it is applied three times (to $PQUQ^TP^T$, $P$ and $P^T$ respectively) when computing $c_{num}(U,P,Q)$ (see Eq.~(\ref{eqn:costf})). While it does impose a significant computational overhead, it gives a better reflection of the true cost function, since the reduced circuit is the circuit that one would use for implementation. Figure \ref{fig:red} shows an example result of applying the reduction procedure to a quantum circuit.

\subsection{MPI parallelisation}
\label{sec:mpiparallel}

The program described above can be readily extended to a parallel architecture using MPI. Since the neighbourhood operator in the simulated annealing procedure acts to interchange any two random positions, it follows that if the random number generator is seeded differently, then a different set of positions would be interchanged, i.e. a different search through the space of permutations would be conducted. Similarly, the qubit permutation that is generated would also change when seeded differently, which enables the program to start threads at multiple locations in the search space of $m!$ permutations, so that the search procedure explores as much of the permutation space as possible. We do, however, restrict the root thread (thread index 0) of the program to use the identity qubit permutation for comparison purposes. Hence, using MPI, we can spawn a team of threads that simultaneously searches through the space of permutations independently and differently (by seeding the random number generator of each thread differently), and then collate the results to pick out the thread with the most optimal permutation, that is, it has the lowest $c_{num}(U,P,Q)$ value.

\section{Results}
\label{sec:res}

We now apply the software program \emph{OptQC} to various unitary operations to obtain corresponding optimised quantum circuits. All the results shown here are obtained using parameters $i_{max} = 40000$, $j_{max} = 1000$ and $\alpha = 0.01$ (we choose this $\alpha$ value because it provides, on average, the best results for the unitary operators being considered in this paper. Using these parameters, we run \emph{OptQC} on the supercomputer Fornax with Intel Xeon X5650 CPUs, managed by iVEC@UWA, using 8 nodes with 12 cores on each (i.e. 96 threads). 

\subsection{Real unitary matrix}

A random real unitary (i.e. orthogonal) matrix is given below: 

\begin{equation*}
U = 
\left(
\begin{array}{cccccccc}
 0.0438 & 0 & 0 & 0 & 0.9990 & 0 & 0 & 0 \\
 0.1297 & 0.8689 & -0.2956 & 0 & -0.0057 & 0.1538 & -0.3423 & 0 \\
 -0.2923 & 0 & 0.6661 & 0 & 0.0128 & 0 & -0.6861 & 0 \\
 -0.0061 & -0.0412 & 0.0140 & 0.7058 & 0.0003 & 0.3008 & 0.0162 & -0.6397 \\
 0.9147 & 0 & 0.4021 & 0 & -0.0401 & 0 & 0 & 0 \\
 0.0185 & 0.1242 & -0.0422 & 0.3961 & -0.0008 & -0.9073 & -0.0489 & 0 \\
 0.2424 & -0.4762 & -0.5524 & 0 & -0.0106 & 0 & -0.6397 & 0 \\
 0.0051 & 0.0343 & -0.0117 & -0.5874 & -0.0002 & -0.2503 & -0.0135 & -0.7686 \\
\end{array}
\right)
\end{equation*}

Note that this matrix is not completely filled, otherwise no reduction via permutations would generally be possible. By using \emph{OptQC}, the reduction process gives the following results for the thread which achieves the optimal solution:

\begin{itemize}
\itemsep0em
\item No optimisation: $c_{num}(U,I,I) = 29$ gates
\item After selection of an optimised qubit permutation $q$: $c_{num}(U,I,Q) = 1+ 26 + 1 = 28$ gates
\item After simulated annealing process for the permutation $p$: $c_{num}(U,P,Q) = 1 + 2+ 16 + 2 + 1 = 22$ gates
\end{itemize}

\begin{figure}[htp]
\centering
\subfigure[Original circuit - 29 gates]{\label{fig:randrealcir1}\includegraphics[scale=0.40]{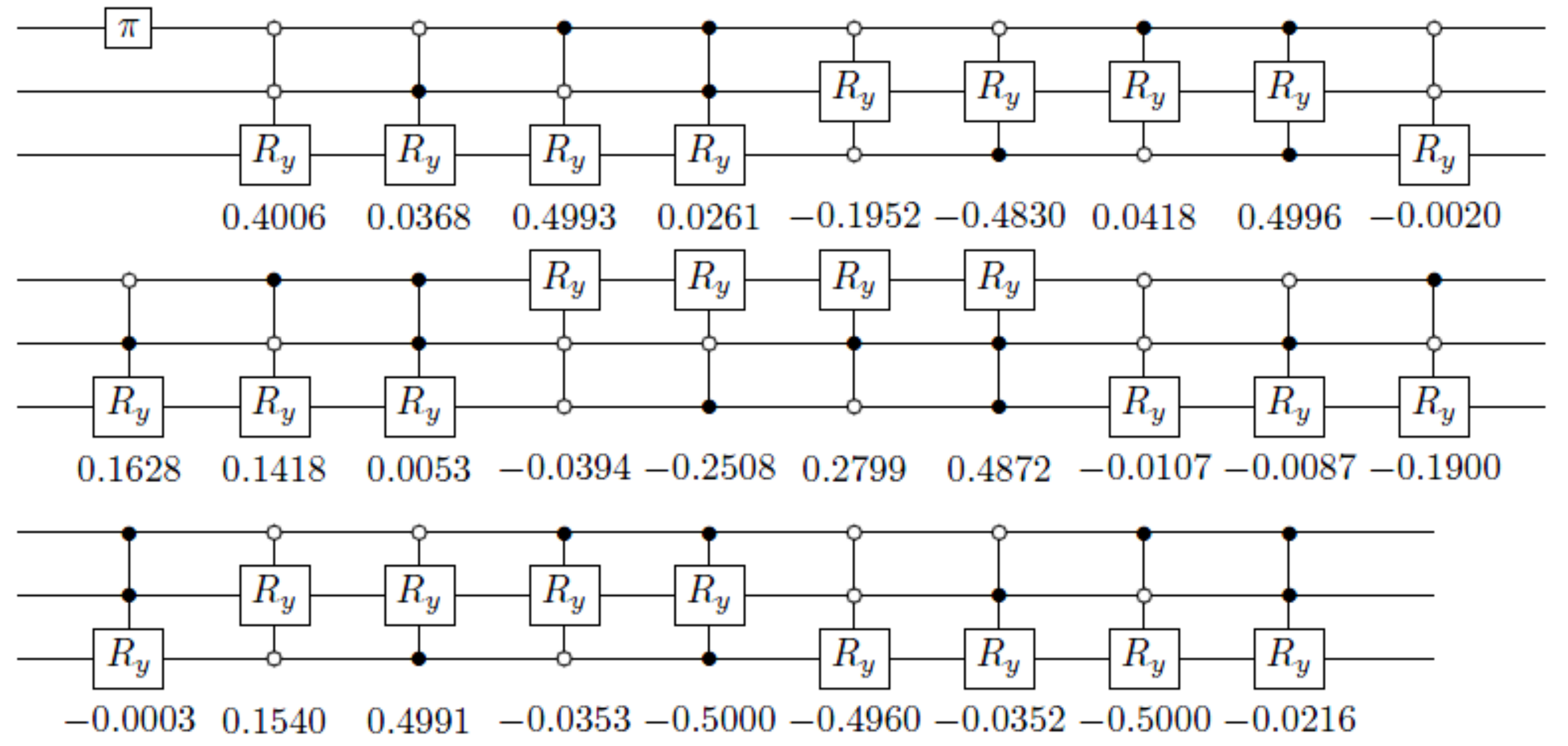}\qquad}
\subfigure[Optimised circuit - 22 gates]{\label{fig:randrealcir2}\includegraphics[scale=0.40]{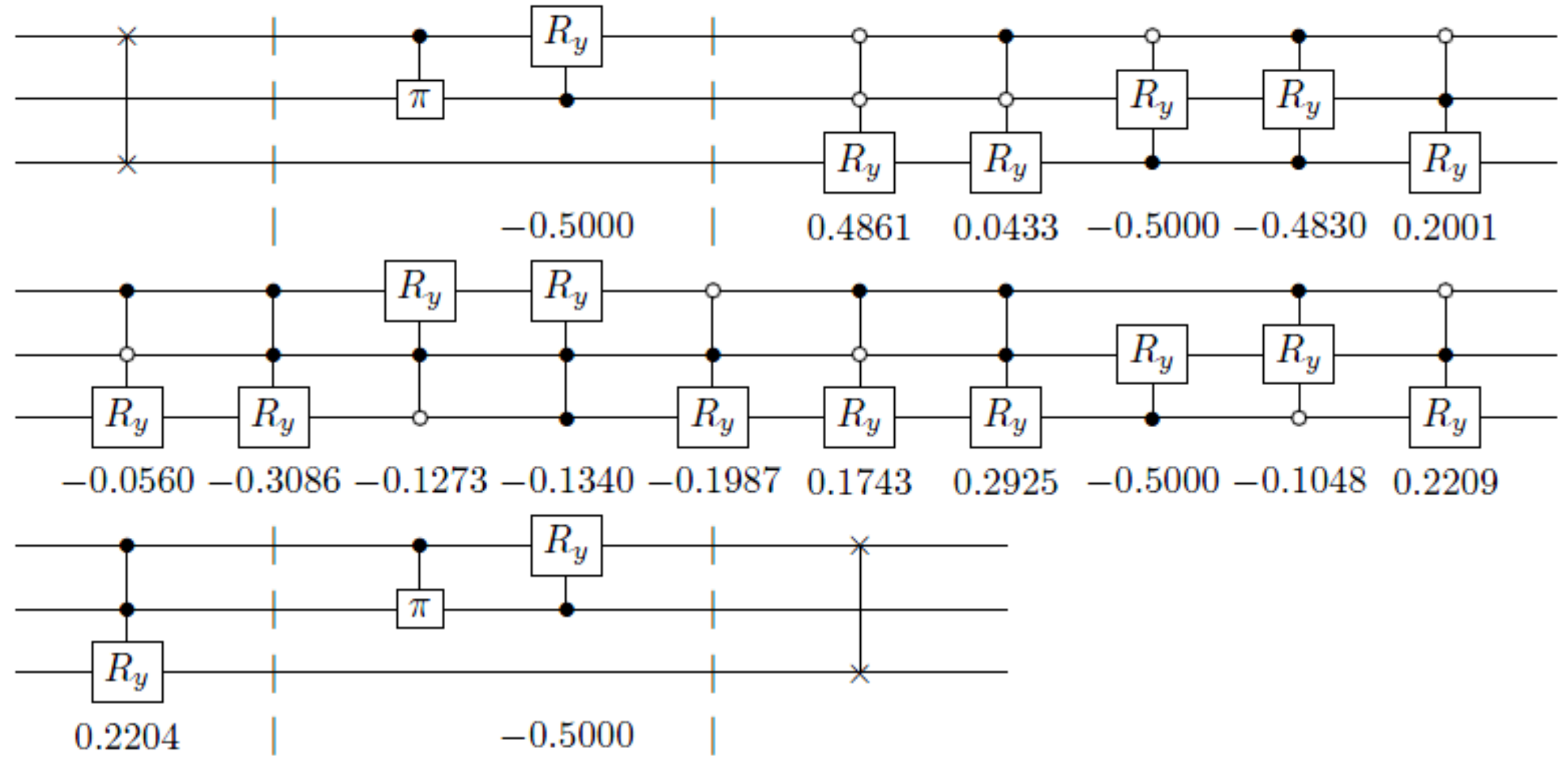}\qquad}
\caption{Result of quantum circuit optimisation as performed by \emph{OptQC} on a random real unitary matrix. In (b), the dashed vertical lines separate the circuit for each matrix - from left to right, this corresponds to $Q$, $P$, $U'$, $P^T$ and $Q^T$ respectively.}
\label{fig:randrealcir}
\end{figure}

Hence, we achieve a reduction of $\sim 25\%$ from the original number of gates. Figure \ref{fig:randrealcir} shows a comparison between the original and optimised circuit for $U$. Runtime for this calculation is $\sim 14.5$ seconds.


\subsection{Quantum walk operators}

One important class of unitary operators are quantum walk operators - in particular, discrete-time quantum walk (DTQW) step operators \cite{kempe_quantum_2003, berry_two-particle_2011, loke_efficient_2012}. For a given undirected graph $ G(V,E) $, defined by a vertex set $V$ and edge set $E$, we can define the DTQW step operator $ U = SC $, where $S$ and $C$ are the shifting and coin operators respectively. The shifting operator acts to swap coin states that are connected by an edge, and the coin operator acts to mix the coin states at each individual vertex. 

\subsubsection{8-star graph}

\begin{figure}[htp]
\centering
\includegraphics[scale=0.30]{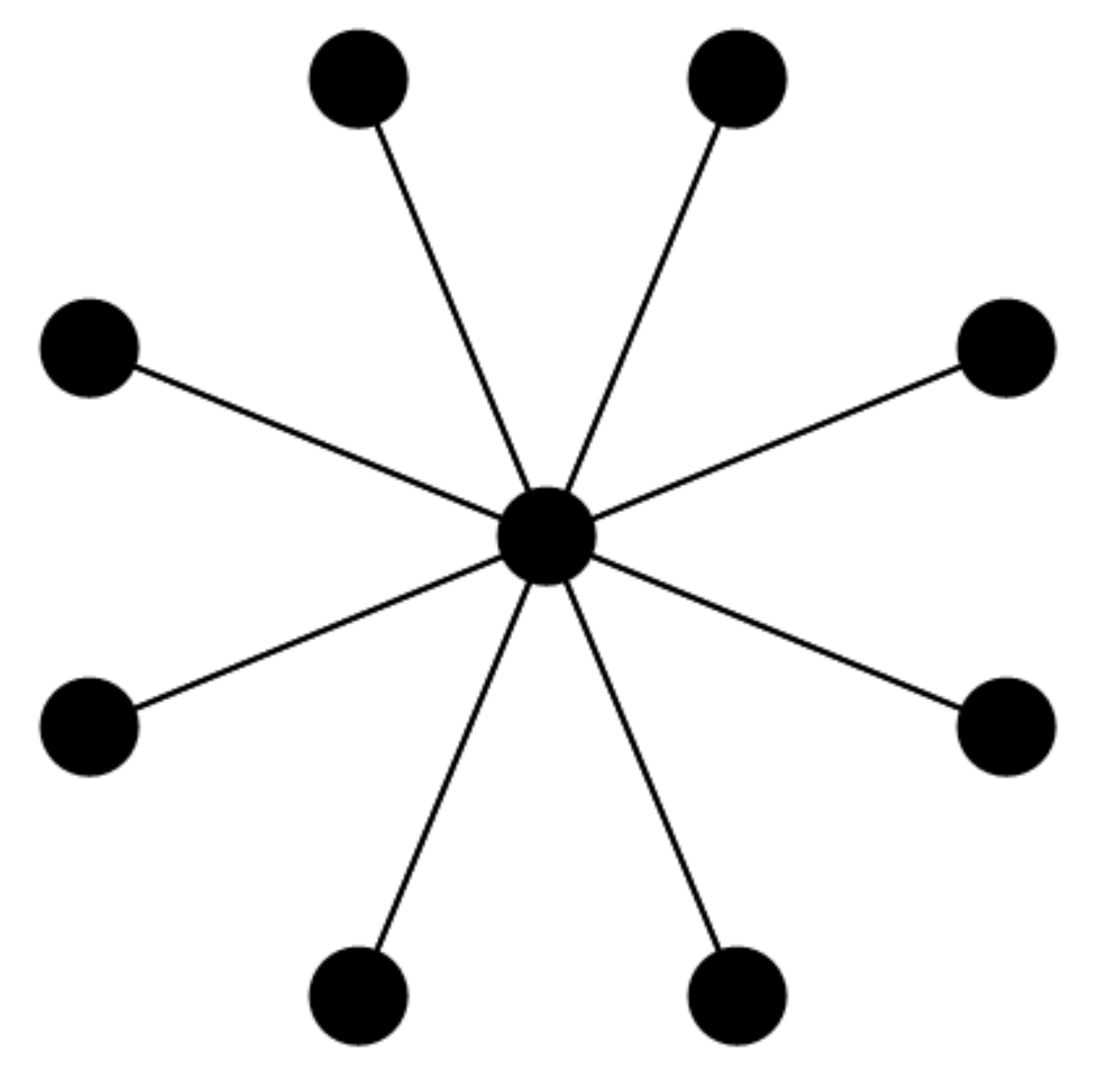}
\caption{The 8-star graph}
\label{fig:S8graph}
\end{figure}

The 8-star graph (shown in Figure \ref{fig:S8graph}) is a graph with 1 centre vertex connected to 8 leaf vertices by undirected edges. Using the Grover coin operator, the resulting quantum walk operator on this graph corresponds to a 16-by-16 real unitary matrix, as given below:

\begin{equation*}
U = 
\left(
\begin{array}{cccccccccccccccc}
 0 & 0 & 0 & 0 & 0 & 0 & 0 & 0 & -0.75 & 0.25 & 0.25 & 0.25 & 0.25 & 0.25 & 0.25 & 0.25 \\
 0 & 0 & 0 & 0 & 0 & 0 & 0 & 0 & 0.25 & -0.75 & 0.25 & 0.25 & 0.25 & 0.25 & 0.25 & 0.25 \\
 0 & 0 & 0 & 0 & 0 & 0 & 0 & 0 & 0.25 & 0.25 & -0.75 & 0.25 & 0.25 & 0.25 & 0.25 & 0.25 \\
 0 & 0 & 0 & 0 & 0 & 0 & 0 & 0 & 0.25 & 0.25 & 0.25 & -0.75 & 0.25 & 0.25 & 0.25 & 0.25 \\
 0 & 0 & 0 & 0 & 0 & 0 & 0 & 0 & 0.25 & 0.25 & 0.25 & 0.25 & -0.75 & 0.25 & 0.25 & 0.25 \\
 0 & 0 & 0 & 0 & 0 & 0 & 0 & 0 & 0.25 & 0.25 & 0.25 & 0.25 & 0.25 & -0.75 & 0.25 & 0.25 \\
 0 & 0 & 0 & 0 & 0 & 0 & 0 & 0 & 0.25 & 0.25 & 0.25 & 0.25 & 0.25 & 0.25 & -0.75 & 0.25 \\
 0 & 0 & 0 & 0 & 0 & 0 & 0 & 0 & 0.25 & 0.25 & 0.25 & 0.25 & 0.25 & 0.25 & 0.25 & -0.75 \\
 1 & 0 & 0 & 0 & 0 & 0 & 0 & 0 & 0 & 0 & 0 & 0 & 0 & 0 & 0 & 0 \\
 0 & 1 & 0 & 0 & 0 & 0 & 0 & 0 & 0 & 0 & 0 & 0 & 0 & 0 & 0 & 0 \\
 0 & 0 & 1 & 0 & 0 & 0 & 0 & 0 & 0 & 0 & 0 & 0 & 0 & 0 & 0 & 0 \\
 0 & 0 & 0 & 1 & 0 & 0 & 0 & 0 & 0 & 0 & 0 & 0 & 0 & 0 & 0 & 0 \\
 0 & 0 & 0 & 0 & 1 & 0 & 0 & 0 & 0 & 0 & 0 & 0 & 0 & 0 & 0 & 0 \\
 0 & 0 & 0 & 0 & 0 & 1 & 0 & 0 & 0 & 0 & 0 & 0 & 0 & 0 & 0 & 0 \\
 0 & 0 & 0 & 0 & 0 & 0 & 1 & 0 & 0 & 0 & 0 & 0 & 0 & 0 & 0 & 0 \\
 0 & 0 & 0 & 0 & 0 & 0 & 0 & 1 & 0 & 0 & 0 & 0 & 0 & 0 & 0 & 0 \\
\end{array}
\right)
\end{equation*}

By using \emph{OptQC}, the reduction process gives the following results for the thread which achieves the optimal solution:

\begin{itemize}
\itemsep0em
\item No optimisation: $c_{num}(U,I,I) = 34$ gates
\item After selection of an optimised qubit permutation: $c_{num}(U,I,Q) = 27$ gates
\item After simulated annealing process to select a permutation $p$: $c_{num}(U,P,Q) = 0 + 2 + 19 + 2 + 0 = 23$ gates
\end{itemize}

Hence, we achieve a reduction of $\sim 32\%$ from the original number of gates. Figure \ref{fig:randrealcir} shows the optimized circuit obtained for $U$.  Runtime for this calculation is $\sim 47$ seconds. 

\begin{figure}[htp]
\centering
\includegraphics[scale=0.30]{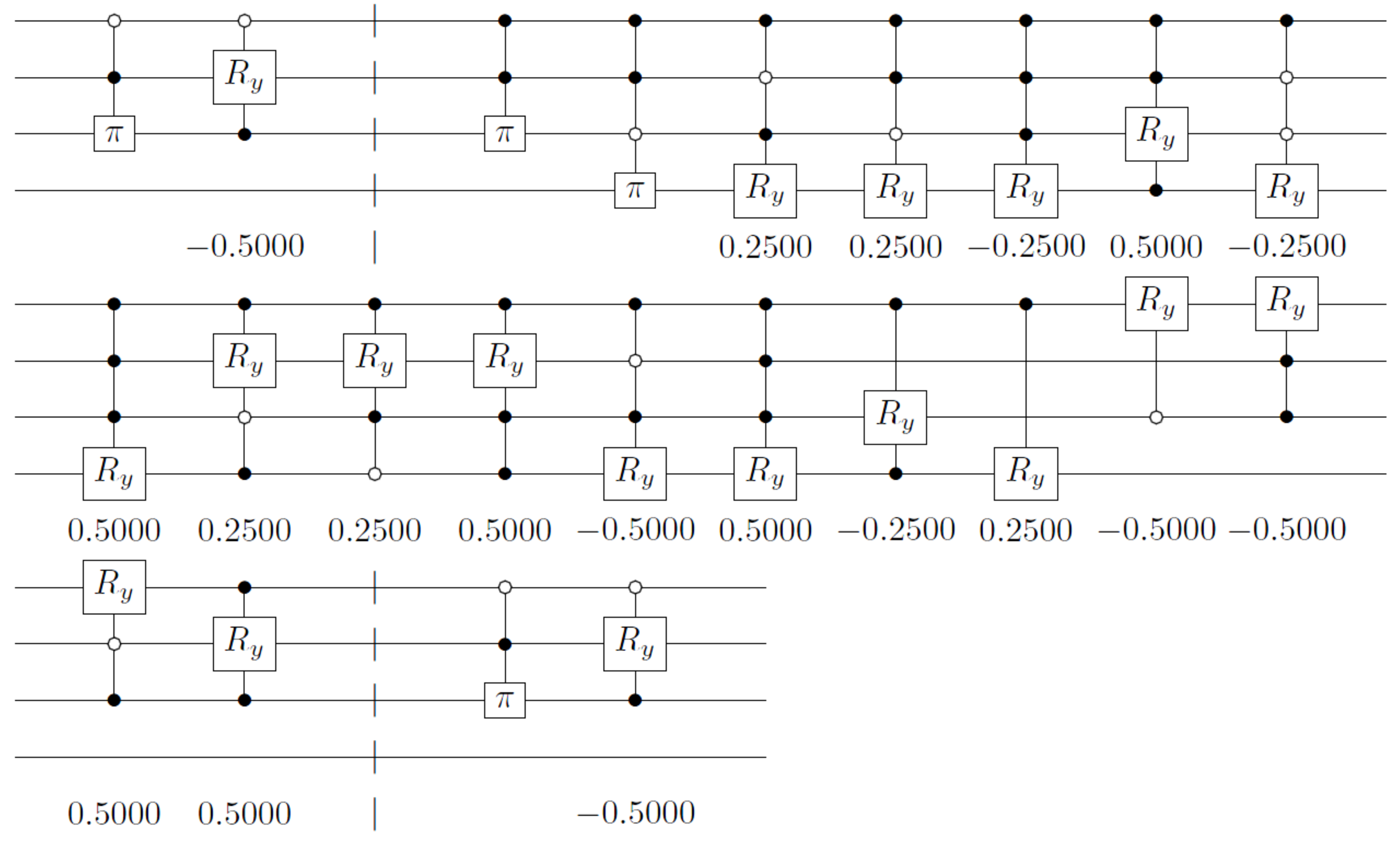}
\caption{Optimised circuit (with 23 gates) for the quantum walk operator of the 8-star graph.}
\label{fig:S8cir}
\end{figure}

\subsubsection{3\textsuperscript{rd} generation 3-Cayley tree}

\begin{figure}[htp]
\centering
\subfigure[3CT3 graph]{\label{fig:3ct3graph}\includegraphics[scale=0.40]{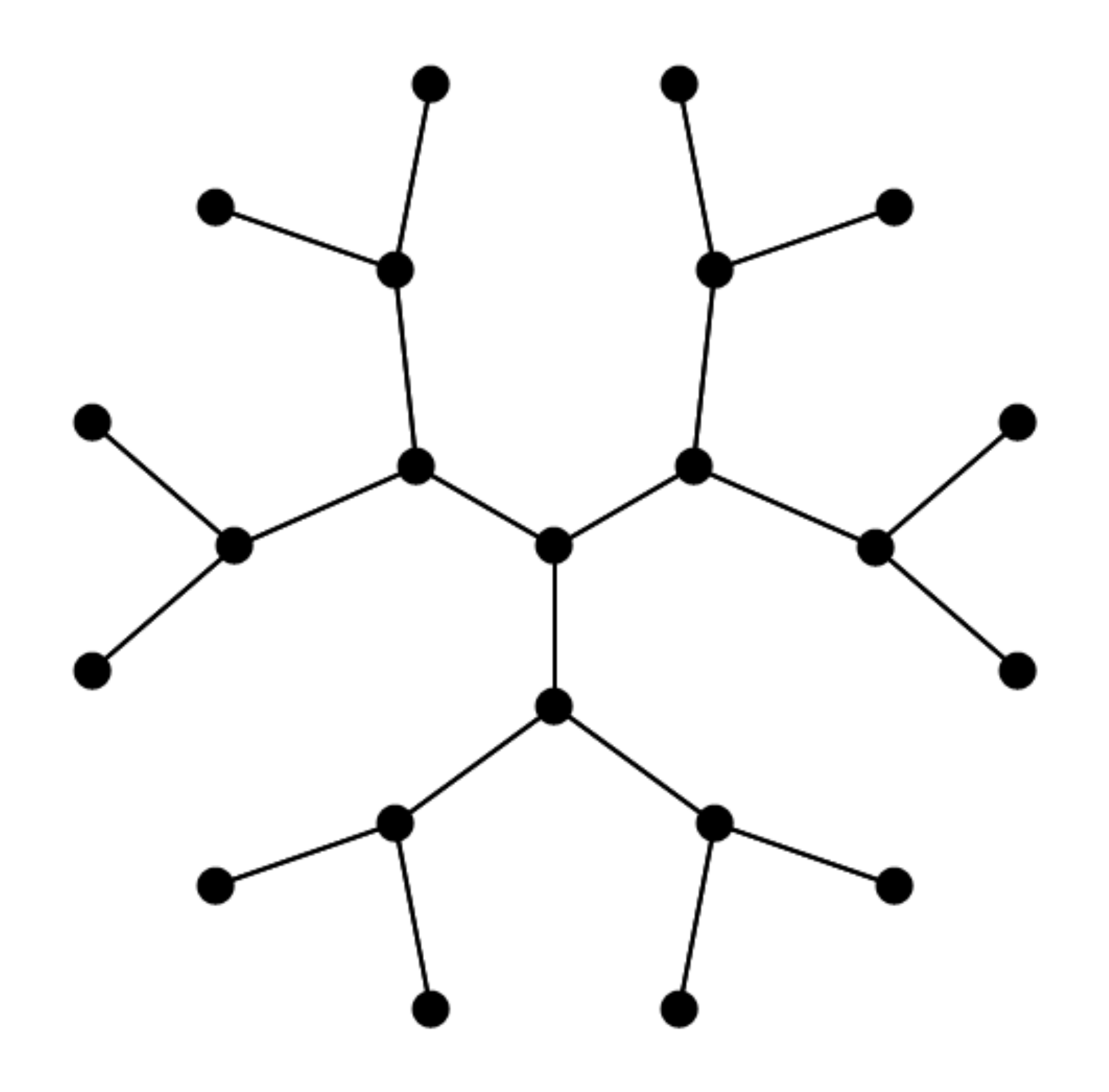}\qquad}
\subfigure[Quantum walk operator for the 3CT3 graph]{\label{fig:3ct3mat}\includegraphics[scale=0.40]{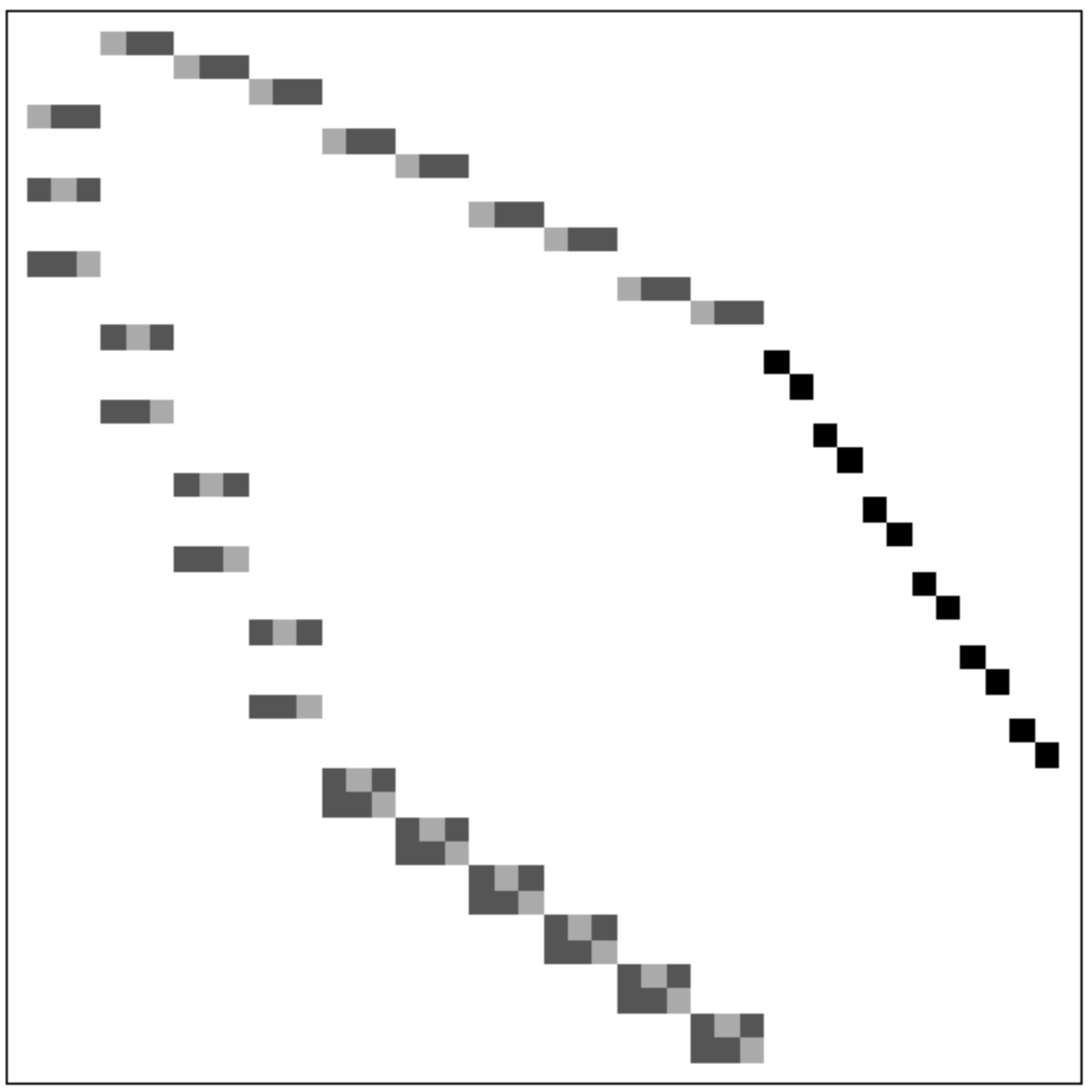}}
\caption{The 3CT3 graph and its corresponding quantum walk operator using the Grover coin operator. The colours/shades in (b) denote the matrix entries for $-1/3$ (light grey), $2/3$ (dark grey) and $1$ (black) - all other matrix entries are $0$ (white).}
\label{fig:3ct3graphmat}
\end{figure}

The 3\textsuperscript{rd} generation 3-Cayley tree (abbreviated as the 3CT3 graph) is a tree of 3 levels in which all interior nodes have degree 3, as shown in Figure \ref{fig:3ct3graph}. The corresponding quantum walk operator using the Grover coin operator is shown in Figure \ref{fig:3ct3mat} - the quantum walk operator $U$ is a 42-by-42 real unitary matrix (which is fairly sparse), which, for the purposes of the decomposition, is expanded to a 64-by-64 unitary matrix as per Eq.~(\ref{eqn:Uexpand}).

By using \emph{OptQC}, the reduction process gives the following results for the thread which achieves the optimal solution:

\begin{itemize}
\itemsep0em
\item No optimisation: $c_{num}(U,I,I) = 996$ gates
\item After selection of an optimised qubit permutation: $c_{num}(U,I,Q) = 3+ 345 + 3 = 351$ gates
\item After simulated annealing process to select a permutation $p$: $c_{num}(U,P,Q) = 3 + 33 + 231 + 30 + 3 = 300$ gates
\end{itemize}

\begin{figure}[htp]
\centering
\includegraphics[scale=0.40]{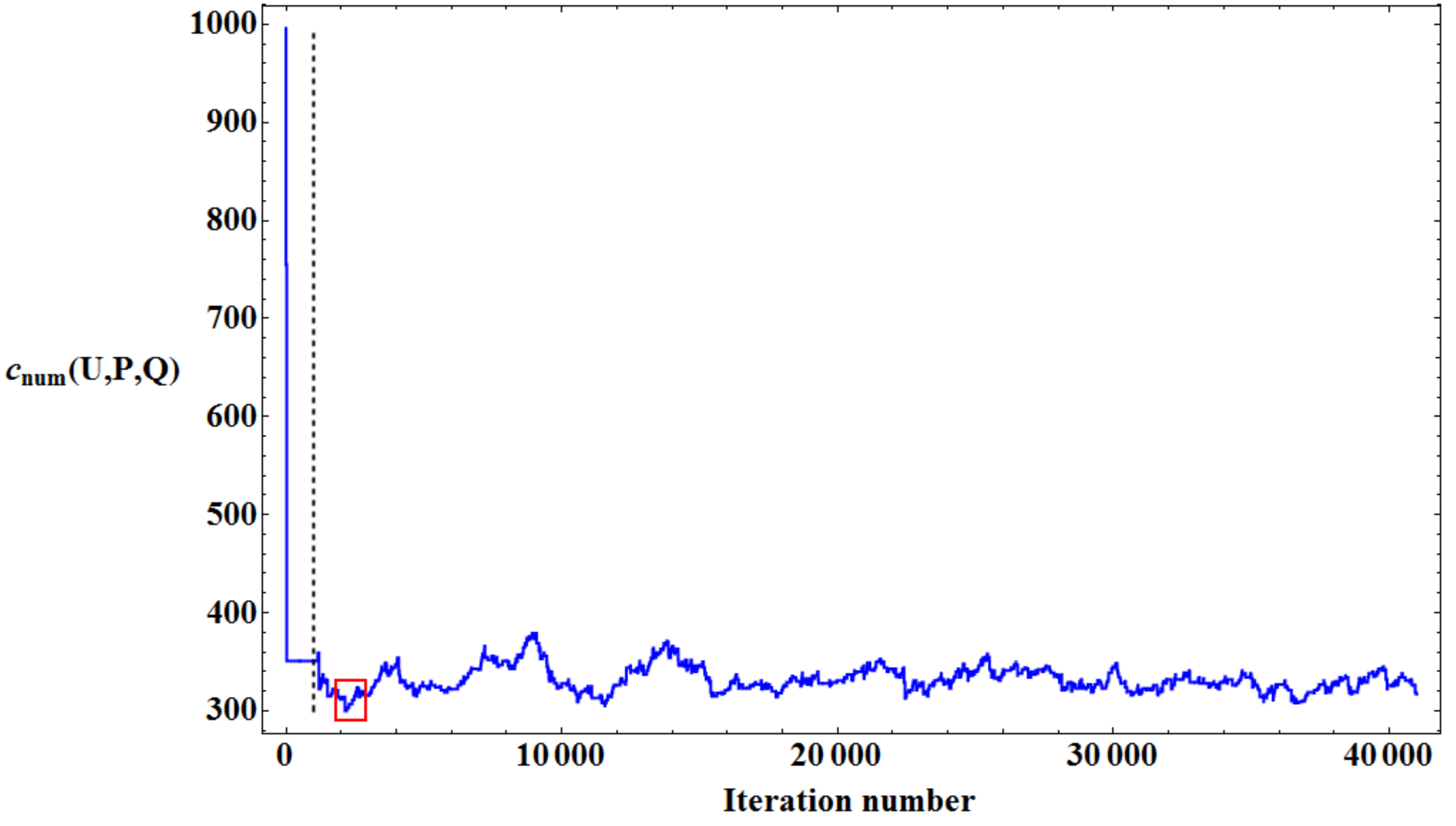}
\caption{Time-series of $c_{num}(U,P,Q)$ during the simulated annealing process for the thread which obtains the optimal solution. The original number of gates required by the CSD method to implement $U$ is 996; selection of a qubit permutation reduces this cost to 351 gates, which is used as the starting point for the simulated annealing process. The red box indicates the region where the optimal solution of 300 gates is achieved. The iterations before the dotted line indicate the qubit permutation selection phase, and the subsequent iterations shows the simulated annealing process.}
\label{fig:3ct3history}
\end{figure}

Hence, we achieve a reduction of $\sim 70\%$ from the original number of gates. Runtime for this calculation is $\sim 12$ minutes. Figure \ref{fig:3ct3history} shows the time-series for $c_{num}(U,P,Q)$ during both the qubit permutation selection phase and the simulated annealing process (separated by a dotted line) to achieve the above result.

\subsection{Quantum Fourier transform}
\label{sec:qftalg}

\begin{figure}[htp]
\centering
\includegraphics[scale=0.20]{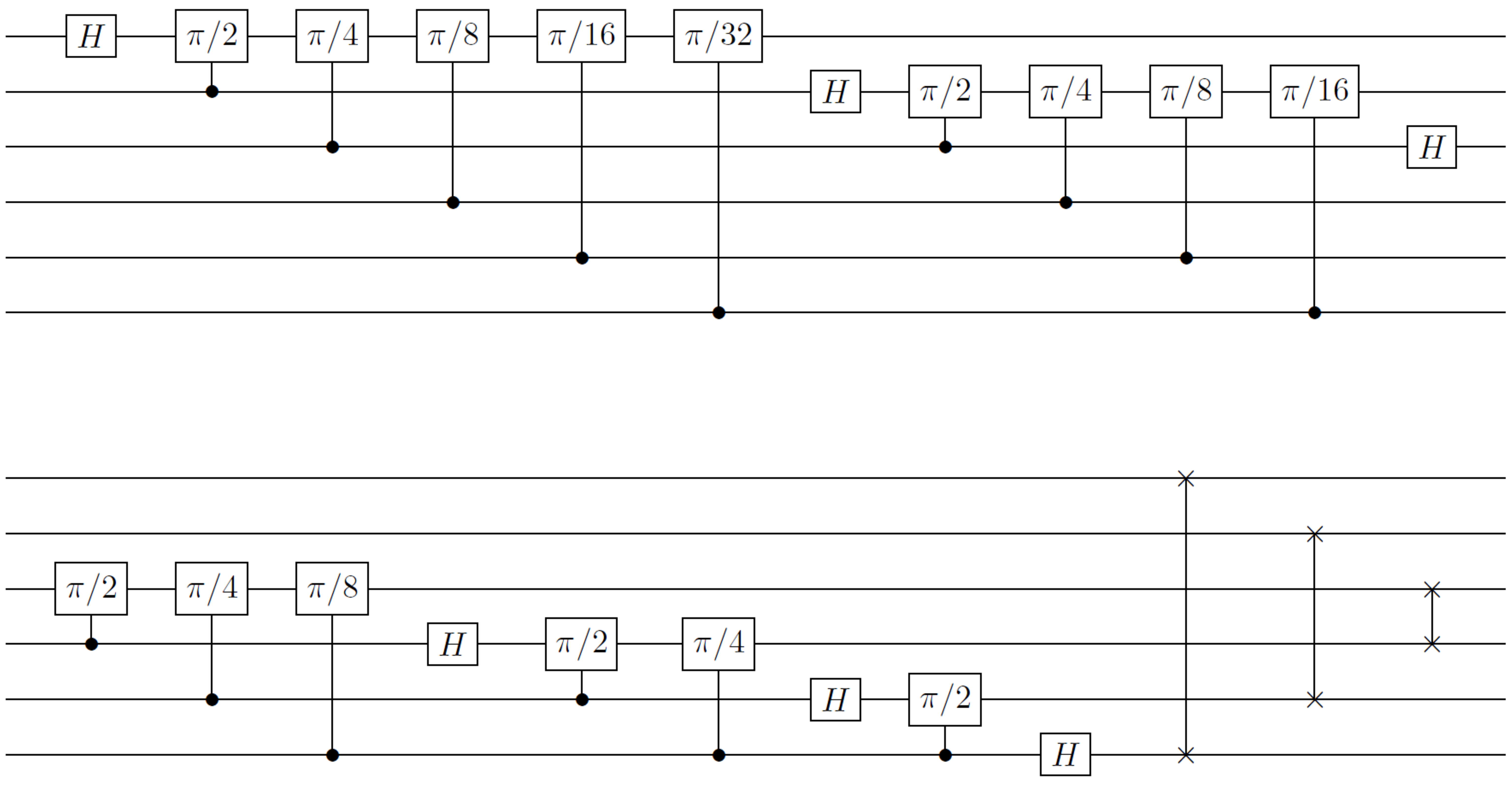}
\caption{Circuit implementation of quantum Fourier transform for $n=64$.}
\label{fig:qft64org}
\end{figure}

Quantum Fourier transform is the quantum counterpart of the discrete Fourier transform in classical computing. It is an essential ingredient in several well-known quantum algorithms, such as Shor's factorization algorithm \cite{shor_polynomial-time_1997} and the quantum phase estimation algorithm \cite{cleve_quantum_1998}. The matrix representation of the quantum Fourier transform on $n$ dimensions is given by:
\begin{equation}
(\mbox{QFT})_{jk} = \frac{1}{\sqrt{n}} \omega^{jk}, \quad \mbox{where  } \omega = \mbox{exp}(2 \pi i / n).
\label{eqn:qftdef}
\end{equation}
An efficient quantum circuit implementation of the quantum Fourier transform is given in \cite{nielsen_quantum_2011}, which scales logarithmically as $O(\mbox{log}(n)^2)$.  Such a circuit implementation for $n=2^6=64$ is shown in Figure \ref{fig:qft64org}.

\begin{figure}[htp]
\centering
\includegraphics[scale=0.30]{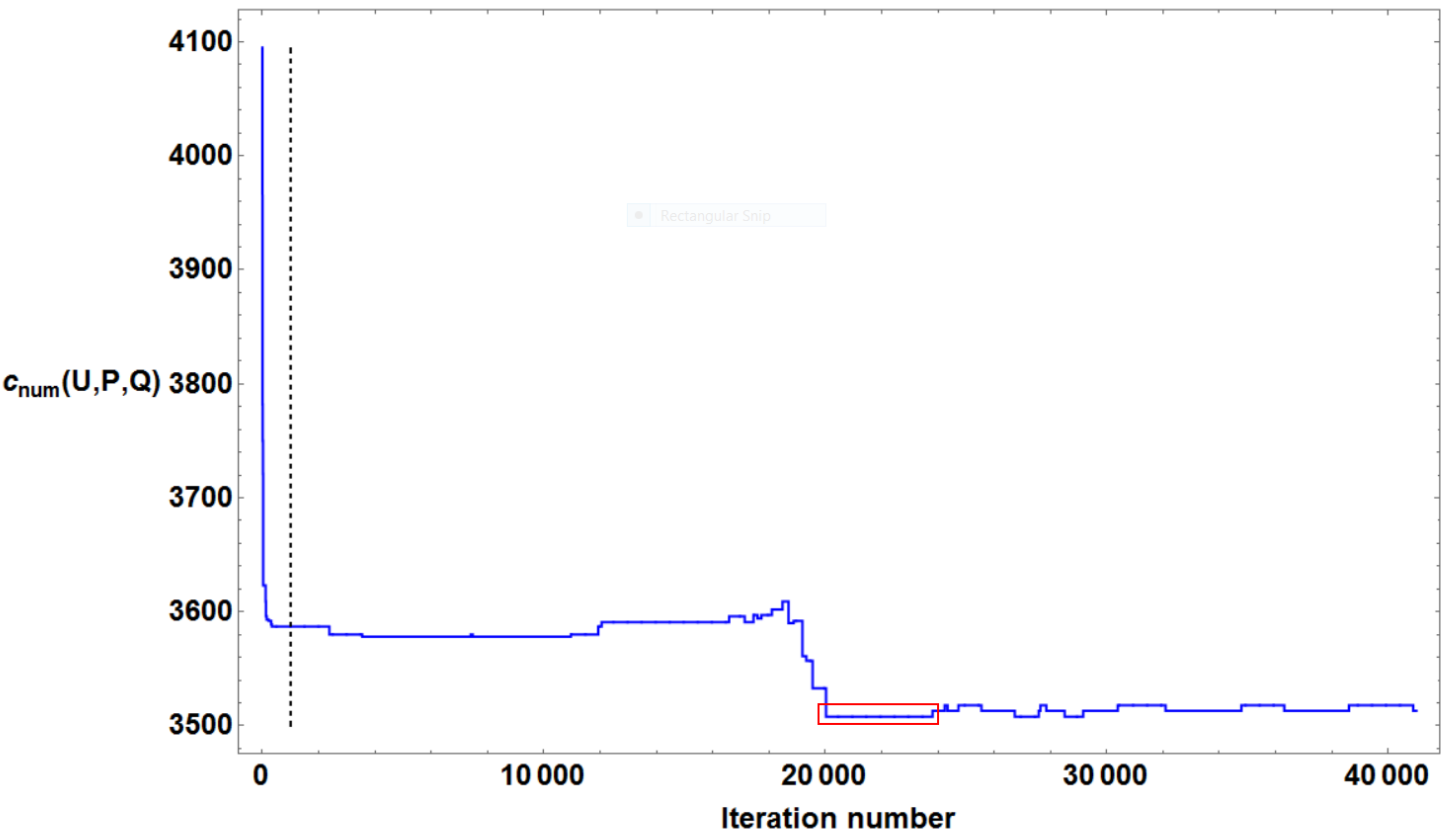}
\caption{Time-series of $c_{num}(U,P,Q)$ during the simulated annealing process for the thread which obtains the optimal solution. The original number of gates required by the CSD method to implement $U$ is 4095; selection of a qubit permutation reduces this cost to 3587 gates, which is used as the starting point for the simulated annealing process. The red box indicates the region where the optimal solution of 3508 gates is achieved. The iterations before the dotted line indicate the qubit permutation selection phase, and the subsequent iterations shows the simulated annealing process.}
\label{fig:qft64history}
\end{figure}

Now, let us apply \emph{OptQC} to the corresponding 64-by-64 complex unitary operator, given by Eq.~(\ref{eqn:qftdef}). With $\alpha = 0.002$, the reduction process gives the following results for the thread achieving an optimal solution:
\begin{itemize}
\itemsep0em
\item No optimisation: $c_{num}(U,I,I) = 4095$ gates
\item After selection of an optimised qubit permutation: $c_{num}(U,I,Q) = 5 + 3577 + 5 = 3587$ gates
\item After simulated annealing process to select a permutation $p$: $c_{num}(U,P,Q) = 5 + 69 + 3359 + 70 + 5 = 3508$ gates
\end{itemize}
Hence, we achieve a reduction of $\sim 14\%$ from the original number of gates. Runtime of this calculation is $\sim 20$ minutes. Figure \ref{fig:qft64history} shows the time-series for $c_{num}(U,P,Q)$ during both the qubit permutation selection phase and the simulated annealing process (separated by a dotted line) to achieve the above result. Clearly, this result is by far inferior to the quantum circuit of only 24 gates shown in Figure \ref{fig:qft64org}.  Similarly, the \emph{OptQC} package would not be able to provide quantum circuits as efficient as those presented in \cite{douglas_efficient_2009, loke_efficient_2012} for the implementation of quantum walks on highly symmetric graphs.  This is to be expected, since the CS decomposition is a general technique that decomposes a given unitary into a fixed circuit structure using many conditional gates, with an upper bound of O($4^n$).  This algorithm is performed without foreknowledge or explicitly exploiting the structure of the unitary, which would clearly be crucial in achieving the lowest possible number of gates for a given unitary, as exemplified by the above examples. Instead, the \emph{OptQC} package is designed to work for any arbitrary unitary operator for which we do not already have an efficient quantum circuit implementation of, for example, quantum walk operators on arbitrarily complex graphs.  In such cases, we have demonstrated that the \emph{OptQC} package provides optimised quantum circuits that are far more efficient than the original \emph{Qcompiler}.

\section{Conclusion and future work}
\label{sec:cafw}

We have developed an optimised quantum compiler, named as \emph{OptQC}, that runs on a parallel architecture to minimise the number of gates in the resulting quantum circuit of a unitary matrix $U$.  This is achieved by finding permutation matrices $Q$ and $P$ such that $U = Q^TP^TU'PQ$ requires less total number of gates to be implemented, where the implementation for each matrix is considered separately. Decompositions of unitary matrices is done using the CSD subroutines provided in the LAPACK library \cite{anderson_lapack_1999} and adapted from \emph{Qcompiler} \cite{chen_qcompiler:_2013}. \emph{OptQC} utilises an optimal selection of qubit permutations $Q$, a simulated annealing procedure to find $P$, and a combination of similar gates in order to reduce the total number of gates required as much as possible. We find that for many different types of unitary operators, \emph{OptQC} is able to reduce the number of gates required by a significant amount, but its efficacy does vary depending on the unitary matrix given. In particular, this optimisation procedure works well for sparse unitary matrices. 

For future work, we hope to look at characterising the optimal solutions reached to see if the matrix $U'$ (and the associated permutation $P$) have some common preferential structure that leads to a reduced cost of implementation using the CSD method. Such information could be used to implement a guided search for the optimal solution, rather than using random adjustments of the permutation matrix. We also want to characterise `bad' permutations (that is, permutations with a large cost) and avoid them in the search procedure, perhaps by eliminating the conjugacy class of `bad' permutations from the search space.

\section{Acknowledgements}

Our work was supported through the use of advanced computing resources located at iVEC@UWA, as well as funding for a summer internship by iVEC.  The authors would like to acknowledge valuable discussions with Chris Harris at iVEC@UWA, and also thank the referees for their constructive comments and suggestions. T.L. is supported by the International Postgraduate Research Scholarship, Australian Postgraduate Award and the Bruce and Betty Green Postgraduate Research Top-Up Scholarship.

\clearpage

\bibliographystyle{model1-num-names}

\bibliography{References}

\clearpage

\end{document}